\documentclass[10pt,twocolumn]{IEEEtran}
\pdfoutput=1
\usepackage{cite}
\ifCLASSINFOpdf
   \usepackage[pdftex]{graphicx}
   \DeclareGraphicsExtensions{.pdf,.jpeg,.png}
\else
   \usepackage[dvips]{graphicx}
   \DeclareGraphicsExtensions{.eps}
\fi
\usepackage{amssymb}
\usepackage{upgreek}
\usepackage{latexsym}
\usepackage{multirow}
\usepackage{epstopdf}
\usepackage{algpseudocode}
\usepackage{xcolor}

\usepackage[cmex10]{amsmath}
\usepackage{amssymb}
\usepackage{array}
\usepackage{mdwmath}
\usepackage{mdwtab}
\usepackage{pslatex}
\usepackage{color}
\usepackage{textcomp}
\usepackage{placeins}
\usepackage{eqparbox}
\usepackage{url}
\usepackage{stfloats}
\usepackage{multicol}
\usepackage{subfigure}
\usepackage{amsthm}
\usepackage{comment}
\usepackage{balance}
\usepackage{gensymb}

\begin{document}
	
	\title{Challenges of Multi-Factor Authentication for Securing Advanced IoT~(A-IoT) Applications}
	
	\author{Aleksandr~Ometov, Vitaly~Petrov, Sergey~Bezzateev,\\ Sergey~Andreev, Yevgeni~Koucheryavy, and Mario Gerla
		\thanks{A.~Ometov, V.~Petrov, S.~Andreev, and Y.~Koucheryavy are with Tampere University of Technology, Finland. A.~Ometov and Y.~Koucheryavy are also with National Research University Higher School of Economics, Russia. S.~Bezzateev is with ITMO University, Russia. M.~Gerla is with University of California Los Angeles, USA.}
		\thanks{This work was supported by the Academy of Finland (project PRISMA). 
			The work of the first author was supported by the Nokia Foundation under a personal scholarship.
			The work of the third author was supported by the Academy of Finland under a personal grant.}
		\thanks{A.~Ometov is the contact author: aleksandr.ometov@tut.fi}
		\thanks{\copyright\,\,2019 IEEE. The work has been accepted for publication in IEEE Network, 2019. Personal use of this material is permitted. Permission from IEEE must be obtained for all other uses, including reprinting/republishing this material for advertising or promotional purposes, collecting new collected works for resale or redistribution to servers or lists, or reuse of any copyrighted component of this work in other works.}
	}
	\maketitle
	
	\begin{abstract}
		The unprecedented proliferation of smart devices together with novel communication, computing, and control technologies have paved the way for the Advanced Internet of Things~(A-IoT). This development involves new categories of capable devices, such as high-end wearables, smart vehicles, and consumer drones aiming to enable efficient and collaborative utilization within the Smart~City paradigm. While massive deployments of these objects may enrich people's lives, unauthorized access to the said equipment is potentially dangerous. Hence, highly-secure human authentication mechanisms have to be designed. \textcolor{black}{At the same time, human beings desire comfortable interaction with their owned devices on a daily basis, thus demanding the authentication procedures to be seamless and user-friendly, mindful of the contemporary urban dynamics.} In response to these unique challenges, this work advocates for the adoption of multi-factor authentication for A-IoT, such that multiple heterogeneous methods -- both well-established and emerging -- are combined intelligently to grant or deny access reliably. We thus discuss the pros and cons of various solutions as well as introduce tools to combine the authentication factors, with an emphasis on challenging Smart City environments. We finally outline the open questions to shape future research efforts in this emerging field.
	\end{abstract}
	\IEEEpeerreviewmaketitle

	\section{Introduction and Rationale}
	\label{sec:intro}
	
	The vision of the Internet of Things~(IoT) opens a new era of technology penetration into the human lives, which touches upon a wide range of use cases: from Smart Home to Smart City and from Smart Grid to Factory Automation~\cite{extra1}.     \textcolor{black}{The numbers of IoT devices that can collect, store, combine, and analyze the massive amounts of data around them by producing valuable knowledge and making relevant actions is growing uncontrollably in an attempt to offer decisive societal benefits while handling both routine and critical tasks across multiple~verticals~\cite{cisco2017global}.}
	
	As it simplifies the lives of people, the IoT also brings unprecedented security and privacy risks, since close to any object around us becomes interconnected with others to collect and process sensitive information~\cite{lin2017survey}. The conventional \emph{massive} IoT involves numerous low-cost devices (e.g., sensors, actuators, and smart meters), with limited computational capabilities and stringent power constraints; hence, the traditional security and privacy solutions had to be reconsidered and adjusted to the specifics of massive IoT. \textcolor{black}{Over recent decades, security and privacy in IoT remained a major research topic subject to heated discussions, e.g., in the area of lightweight cryptography~\cite{shamir2017summary}, secure connection and trust establishment~\cite{guo2017survey}, and privacy-preserving data processing~\cite{zhou2017security}.} While there are multiple open problems yet to be resolved, the current progress in this field promises to provide the demanded levels of security to these massive IoT~deployments.
	
\begin{figure}[!ht]
		\centering
		\includegraphics[width=0.8\columnwidth]{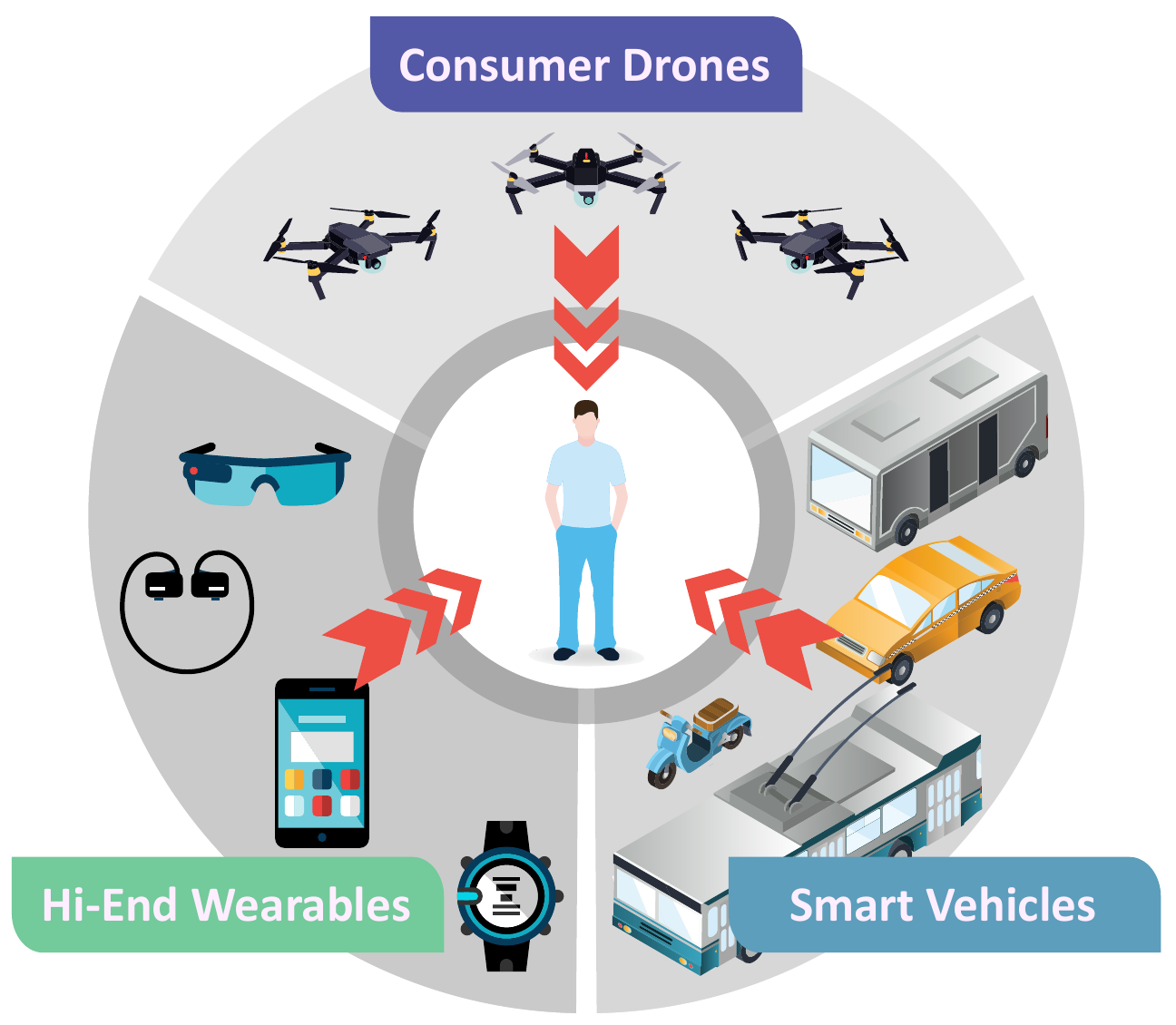}
		\caption{Human-centric Advanced IoT (A-IoT) applications in a Smart City.}
		\label{fig:human}
	\end{figure}

	Meanwhile, in contrast to the massive and low-cost IoT solutions, an emerging trend in today's IoT is a rapid proliferation of high-end IoT equipment that features more capable connected devices. These include sophisticated wearables (including augmented, virtual, and mixed reality systems), smart vehicles, and consumer drones (see Fig.~\ref{fig:human}) -- that may collectively be named \emph{Advanced IoT~(A-IoT)}. These relatively high-cost devices have more abundant performance, memory, and battery resources to execute full-scale security and privacy protocols; thus, the establishment of secure machine-to-machine connections may not be a challenging problem for the~A-IoT.
	
	At the same time, a number of specific security and privacy concerns emerge in connection with such systems. Since unauthorized access to these powerful devices may lead to severe risks that range from theft of this high-cost equipment (drones or cars) and up to putting human lives in danger by, e.g., manipulating with the information projected to the augmented reality grasses or maneuvering smart vehicles uncontrollably~\cite{joy2018internet}. \textcolor{black}{Therefore, reliable assessment of the \textit{fact of ownership} for the A-IoT devices that belong to both personal and collective use becomes one of the critical challenges that is faced today, which is very different from massive~IoT.}
	
	In this work, we first systematically review the unprecedented research challenges related to determining the human ownership of the A-IoT systems. We then classify the specific features of the A-IoT that can be employed to securely verify the fact of ownership of the A-IoT devices and map them onto the challenges by illustrating how the features can complement each other while covering the potential issues. We also discuss the concept of multi-factor human authentication with the A-IoT system, where multiple heterogeneous factors are intelligently combined to achieve higher levels of security while not compromising the usability of the A-IoT services. We finally enumerate the important practical matters to be resolved on the way towards successful implementation of the introduced concept.
	
	\section{Challenges of Determining Ownership in A-IoT}
	\label{sec:limitations}
	
	\textcolor{black}{As unauthorized access to A-IoT systems brings severe security threats, the challenge of reliable access control becomes one of the most crucial research problems for securing A-IoT solutions.} An access control procedure can generally be decomposed into user authentication and authorization.     \textcolor{black}{The second stage is relatively less complicated and can be implemented by conventional discretionary, mandatory, or role-based access control methods.} However, the first stage introduces a number of A-IoT-specific research questions that we carefully review in this section.
	
	\subsection{Multi-Modality of Human-Computer Interaction}
	Today, most of the conventional ICT systems are equipped with advanced input devices, such as keyboards and touchscreens, as well as output devices, most commonly, LCD screens used for human-computer interaction~(HCI). \textcolor{black}{Since textual input remains the dominating form of HCI, these systems have historically been adopted for authentication purposes: memorable textual or numerical passwords, possibility to display a hint or advanced visual instructions, etc.}
	
	In contrast, the very nature of authentication does not imply text-based commands or responses. Very few of the emerging IoT devices are controlled by a keyboard; hence, the authentication methods based on textual passwords will need to evolve accordingly for them to continue being usable on the mass IoT market.
	
	\subsection{Robustness to Environment and User Behavior}
	The authentication process of today is typically applied in dedicated, comfortable, and stationary environments. Many such actions occur indoors, where neither weather conditions nor other unpredictable factors can impact the authentication decisions. Even when this process happens outdoors, the input devices to enter the security credentials acquire additional protection to resist the environmental changes up to a certain~extent.
	
	However, the A-IoT systems in Smart Cities are mobile by design. Their interactions with a user are spontaneous and occur in uncontrolled and unpredictable environments. Moreover, even under regular weather/environment conditions, the initial state as the user begins interacting with the A-IoT system may be notably different. \textcolor{black}{For example, the user opening a vehicle may be wearing gloves during winter time, such that a fingerprint scanner installed on the door handle may not be available.} Therefore, authentication of A-IoT devices must be made robust to both dynamic environmental conditions and flexible user behavior.
	
	\subsection{High Levels of Reliance and Trust}
	\textcolor{black}{Broad penetration of ICT systems on the consumer market and their role in the daily human life have always been associated with a level of trust that people grant to these systems.} High trust is impossible to achieve without appropriate authentication and authorization procedures~\cite{pascal}. 
	
	At the same time, the A-IoT systems are more elaborate than the ICT platforms of today. They are often granted direct access to sensitive personal information; hence, the data they collect and handle should not be made available to potential third parties. On the other hand, large vehicles, drones, and industrial robots represent more capable platforms, sometimes termed as \emph{sources of increased danger}. This recognizes that they may become hazardous as long as health and even lives of humans are concerned. Therefore, A-IoT systems have to be featured with more secure and reliable authentication procedures, so that they are capable of distinguishing their valid user from an unauthorized adversary.
	
	\subsection{Constrained Response Times and Usability}
	Regardless of their stringent security demands, the response levels of A-IoT authentication are also crucial for its successful adoption. \textcolor{black}{Previously, authentication process was a dedicated phase of the HCI, thus making users prepare for it both physically and mentally: recall the secret phrase, bring the token key, etc.} With further development and penetration of the A-IoT systems, they become more ubiquitous and omnipresent. In future Smart Cities, users will be interacting with various A-IoT devices numerous times a day; hence, they cannot afford to spend several second by authenticating with each of those and tolerate second-long delays in acquiring access.
	
	In response to these demands, A-IoT authentication must evolve to become capable of operating within stringent time intervals, preferably in an inconspicuous form, i.e., transparent to the user. For multi-functional A-IoT systems, this may even bring the need to temporarily provide access to certain basic functionality sooner, while more rigorous authentication is performed in the background. This is because the users are unlikely to require sensitive actions from the very first moments of their interaction with the target A-IoT platform.
	
	
	
	From the above, it follows that designing adequate A-IoT authentication mechanisms is challenging. However, the more advanced capabilities and functions of A-IoT devices can be beneficial when coining novel authentication schemes and we review these in the next section.
	
	\section{Enablers for Improved A-IoT Authentication}
	\label{sec:enablers}
	
	\textcolor{black}{Reliable human user authentication by the A-IoT system is a complex task due to various challenges as discussed previously.} Fortunately, modern A-IoT platforms feature a number of dedicated input devices as well as rich sensing, communication, and computation capabilities, which altogether can be employed during the authentication stage. \textcolor{black}{Various user authentication methods become suitable for new A-IoT systems utilizing this diverse functionality.} In this section, we discuss these authentication methods and their applicability in the A-IoT systems. For convenience, we sort them by following their mass adoption: from well-known to emerging, see Table~\ref{tab:comp}.
	
	\begin{table}[!h]
		\caption{Authentication factors suitable for A-IoT. 
			Type:~K~--~knowledge;~O~--~ownership; BI -- biometric; BE~--~behavior.
			Action:~A~--~active;~P~--~passive.
			Duration:~S~--~short~($<1$~sec);~M~--~medium ($1-15$ sec); L~--~Long~($>15$~sec).}\label{tab:comp}
		\centering
		\def\arraystretch{1.2}
		\begin{tabular}{m{4cm} ccc}
			\hline\hline
			\textbf{Factor} & \textbf{Type} & \textbf{Action} & \textbf{Duration} \\  \hline
			\hline
			\textbf{PIN code}                                     & K     & A     & S     \\\hline
			\textbf{Password}                                     & K     & A     & M     \\\hline
			\textbf{Token}                                         & O     & P     & S     \\\hline
			\textbf{Voice}                                         & BI/BE & A/P     & S/M     \\\hline
			\textbf{Facial}                                     & BI     & A/P     & S/M     \\\hline
			\textbf{Ocular-based}                                 & BI     & A     & S/M     \\\hline
			\textbf{Fingerprint}                                 & BI     & A/P     & S     \\\hline
			\textbf{Hand geometry}                                 & BI     & A/P     & S     \\\hline
			\textbf{Geographical location}                         & BE     & P     & L     \\\hline
			\textbf{Vein recognition}                             & BI     & A/P    & S     \\\hline
			\textbf{Thermal image}                                 & BI/BE & P     & S/M     \\\hline
			\textbf{Behavior patterns}                            & BE     & P     & L     \\\hline
			\textbf{Weight}                                         & BI     & P     & S     \\\hline
			\textbf{Electrocardiographic~(ECG) recognition}     & BI/BE    & P     & S-L     \\\hline
			\hline
		\end{tabular}
	\end{table}
	
	\subsection{Review of Possible Enablers}
	\label{sec:review}
	
	\subsubsection{Hardware tokens}
	\textcolor{black}{The automotive cluster has its legacy security mechanisms, primarily centered around the use of hardware tokens that represent the \textit{ownership} factor.} Recently, such tokens have been complemented by increasingly popular software-based replacements installed on smartphones\footnote{F.~Lardinois, ``BMW~wants~to~turn~your smartphone~into~your~car~key,'' \url{https://techcrunch.com/2018/02/26/bmw-wants-to-turn-your-smartphone-into-your-car-key/}  [Accessed~November~2018]}. By leveraging this concept, the A-IoT systems can make a step forward and utilize the tokens placed not only in the smartphones but also on wearable devices.
	
	\subsubsection{Memorable passwords/PINs}    
	\textcolor{black}{Utilization of conventional PINs is currently acceptable worldwide owing to the widespread adoption of ATMs and early-mobile phone era.} A combination of button presses to unlock a feature (e.g., engine start) or to access a restricted area in addition to the key are typical solutions. Finally, knowledge-based approaches are used widely to access a web-service. The A-IoT systems may intelligently utilize similar solutions as well, where password inputs can effectively become replaced by the use of touchscreen (where applicable) or, e.g., audio forms of input.
	
	\subsubsection{Fingerprint/palm/eye scanner}
	While core technology principles for fingerprint and palm recognition have been known for already a while, the recent achievements in the respective miniaturization made them accessible by a wide range of consumer products, namely, smartphones. Installation of biometric scanning devices within a conventional input interface (e.g., Home button in Apple iPhones) or behind a touchscreen is not a science fiction anymore\footnote{V.~Savov, ``I~tried~the~first~phone~with an~in-display~fingerprint~sensor,'' \url{https://www.theverge.com/circuitbreaker/2018/1/9/16867536/vivo-fingerprint-reader-integrated-display-biometric-ces-2018} [Accessed~November~2018]}. \textcolor{black}{Hence, the authentication process can become transparent for the user, thus improving the overall system usability.} 
	
	\subsubsection{Facial recognition}
	The methods of facial recognition by built-in video cameras originally started with landmark picture analysis, which appeared to be vulnerable to trivial attacks of, e.g., presenting a photo instead of the real face. Over the last two decades, these tools have significantly developed towards three-dimensional face and expression recognition that is much more resilient to such attacks. \textcolor{black}{The security levels can be enhanced further by prompting the user to move the head in a specific manner so that a particular pattern to follow is not known in advance~\cite{corneanu2016survey}.} Solving this task from another angle, a drone can fly around the user to construct a 3D map of face/body without making the user move.
	
	\subsubsection{Voice recognition}
	All of the considered A-IoT devices are typically equipped with a microphone that enables voice recognition. The recently announced implementations are capable of distinguishing millions of different voices after capturing only a short phrase. \textcolor{black}{These solutions are however more vulnerable to a `spoofing attack' than facial recognition. The attack itself could be described as Eve intercepting or capturing digital or analog Alice's voice signal and ``spoofing'' own message into the authentication sensor in the real time (using artificial, synthesized voice) or replicating it later on.} While it is technically possible for an adversary to construct a phrase based on the recorded pronunciation of syllables and sounds, the A-IoT systems are likely to have sufficient computational power for timely recognition of the corresponding attacks.
	
	\subsubsection{Data from wearables}
	\textcolor{black}{The A-IoT devices may also employ their advanced communication capabilities. Particularly, if the authenticating user holds wearable devices, they could act as providers of the authentication factors~\cite{ometov2018multi}, such as gesture analysis, Electrocardiographic~(ECG) Recognition, Geographical Location analysis, etc.} Being connected to the A-IoT system via a short-range radio, wearables can present the security credentials of their user, such as heart rate or electrocardiogram. The utilization of this method requires support from appropriate security protocols, so that the platform may trust the data collected by the user-controlled equipment on the one hand, and the users can be certain that their sensitive personal information is not disclosed, on the~other.
	
	\subsubsection{Behavioral patterns}
	The A-IoT system can utilize one or several input interfaces to record and analyze the individual features of user behavior: response time to typical requests, typing rhythm, micro- or macro-scale mobility, etc. Here, the choice of particular factors to monitor highly depends on the form-factor of the A-IoT device: for wearable electronics these could be accelerometer fingerprinting, for drones they are the control operations, while for smart vehicles there are plenty of options that range from brake pressure and position of hands on the wheel to musical and radio preferences.
	
	\subsection{Mapping Enablers onto Challenges}
	
	While each of the A-IoT-specific authentication methods can bring its additional benefits, none of them alone is capable of efficiently resolving all of the discussed A-IoT challenges. To this end, Table~\ref{tab:mapping} offers a mapping of the authentication methods onto the challenges introduced in Section~\ref{sec:limitations}.
	
	\begin{table*}
		\caption{Comparing A-IoT authentication methods}
		\label{tab:mapping}
		\centering
		\def\arraystretch{1.2}
		\begin{tabular}{m{2.5cm}|c|c|c|c|c}
			\hline\hline
			\textbf{Authentication method} & \textbf{Non-text input} & \textbf{Short contact time} & \textbf{Stringent usability} & \textbf{Environmental robustness} & \textbf{High security level}\\ 
			\hline
			\hline
			Hardware tokens & + & + & - & + & - \\
			\hline
			Password/PIN & - & + & - & + & -\\
			\hline
			Fingerprint/Palm scanner & + & + & +/- & - & +\\
			\hline
			Facial recognition & + & - & + & - & +\\
			\hline
			Voice recognition & + & - & +/- & + & +/-\\
			\hline
			Data from wearables & + & + & - & - & +\\
			\hline
			Behavior patterns & + & - & + & - & +\\
			\hline\hline
		\end{tabular}
	\end{table*}
	
	Notably, knowledge-based methods have their most severe limitations with usability and security requirements~\cite{katsini2016security}, since the user is expected to create, remember, and timely update the secret passwords for all A-IoT devices. In this case, it is very likely that the same password will be selected for multiple systems, which degrades the levels of security. In contrast, hardware tokens are more scalable to be used for multiple A-IoT systems. However, the security levels may still be insufficient as the token(s) can easily be stolen.
	
	\textcolor{black}{Biometrics allow to be authenticated without an additional device or knowledge, but the fingerprint, ocular scanning, or voice recognition may require further effort from the user (e.g., remove gloves or glasses, say a particular phrase, etc.) as well as remain not fully robust to the environmental conditions.} Finally, the risk of losing a biometric template has to be considered. Then, authentication with wearable data has a significant advantage over the conventional voice/face recognition, since the user is not required to perform any explicit action. \textcolor{black}{Meanwhile, this method has similar drawbacks as do the tokens, where the user has to carry the necessary devices continuously, always turned on and charged.}
	
	\textcolor{black}{The methods of behavior recognition allow for mitigating most of the constraints by observing the user behavior over a certain period.} However, the amounts of time necessary for such monitoring are at least an order of magnitude higher than those for other methods, which may become a severe usability concern in delay-sensitive A-IoT applications.     \textcolor{black}{Furthermore, behavior recognition is a complex task from the algorithm design perspective, as there should be a constructive differentiation between a valid deviation in the monitored factor by the actual user and invalid patterns by adversaries.}
	
	As can be concluded from our analysis and Table~\ref{tab:mapping}, neither of the presented methods alone is sufficient to effectively authenticate the user over a broad range of possible scenarios related to the A-IoT systems. In the following section, we propose a novel approach to construct reliable authentication solutions for A-IoT devices by intelligently combining multiple potentially unreliable methods, which follows the multi-factor authentication~(MFA) paradigm.
	
	\begin{figure}
		\centering
		\includegraphics[width=1\columnwidth]{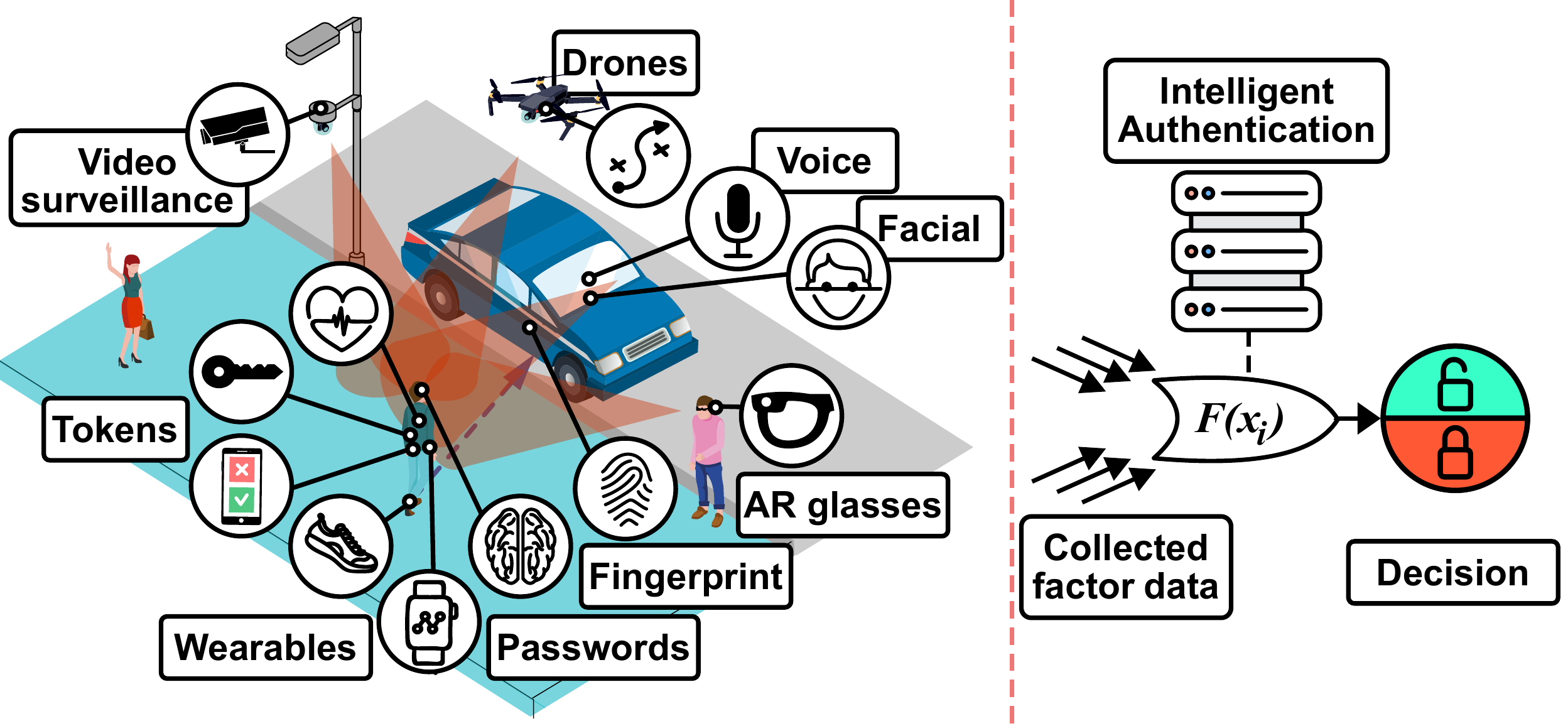}
		\caption{Heterogeneous MFA for A-IoT (by example of smart vehicles).}
		\label{fig:scenario}
	\end{figure}
	 
	
	\section{Use of Multi-Factor Authentication for A-IoT}
	\label{sec:mfa}
	
	Since no single authentication method is likely to be suitable to resolve all of A-IoT challenges, the use of MFA is a natural approach to construct compound solutions (see Fig.~\ref{fig:scenario}).     \textcolor{black}{At the same time, designing adequate MFA mechanisms is a complex matter, which calls for careful selection, harmonization, and combination of various individual methods, such that the resulting solution could outperform its component elements concerning both security and usability, as confirmed by experiments in, for example,~\cite{benaliouche2014comparative}.}     \textcolor{black}{Below, we summarize the four fundamental design principles to be considered when building A-IoT-ready MFA solutions.}
	
	\subsection{Means to Compare}
	
	\textcolor{black}{Before combining several heterogeneous authentication methods, one needs to harmonize across them, such that knowledge-based methods could be integrated with, e.g., biometric and ownership schemes within a single-stop A-IoT authentication mechanism. Importantly, the output of the overwhelming majority of individual authentication solutions is binary: either acceptance or rejection, i.e., $\{0;1\}$. In rare cases, a continuous variable that characterizes the ``likelihood'' ($[0;1]$) could be retrieved from certain biometric systems.} However, most vendors do not provide with access to those values but rather convert the likelihood factor into a binary decision~internally.
	
	\textcolor{black}{In addition to the output data format, alternative methods can be characterized by their accuracy, which is typically estimated with two probabilities: (i)~false acceptance rate~(FAR), the probability that an unauthorized user is accepted; and (ii)~false rejection rate~(FRR), the probability that a valid user is rejected.} These reflect two major qualities of an authentication system: security (FAR) and usability (FRR). 
	We here advocate their generalization to knowledge and ownership~methods. 
	
	For instance, in password-based protection, FAR may correspond to the probability of guessing the secret, while FRR may characterize the possibility of making an accidental mistake during input. In turn, FAR and FRR may also reflect the chances for a token to be stolen or lost for ownership factors. Therefore, we conclude that all of the discussed authentication methods can be well-represented in a unified output format and supplemented with their suitable FAR/FRR values.
	
	\subsection{Means to Combine}
	
	\textcolor{black}{The use of several individual authentication methods does not offer immediate advantages, since it still remains unclear how to combine them efficiently. At the first glance, one may come up with either of the two extreme strategies: ``A~user should successfully pass ALL the checks to receive access'' (\emph{\underline{All}}) and ``A~user should successfully pass ANY of the checks to receive access'' (\emph{\underline{Any}}).}
	
	Below, we present a typical example that numerically illustrates the inherent weaknesses of these extreme strategies as well as emphasizes the importance of a certain level of intelligence when deriving the resulting decision from a number of individual outcomes by the component methods.     \textcolor{black}{We assume a number of factors, each characterized by its own FAR and FRR values.}     \textcolor{black}{For simplicity, we require that all the FARs are equal to~$0.03\%$, whereas all the FRRs are equal to~$2\%$. The Law of Total~Probability then derives the resultant values for FAR/FRR.}
	
\begin{figure}
		\centering
		\includegraphics[width=0.95\columnwidth]{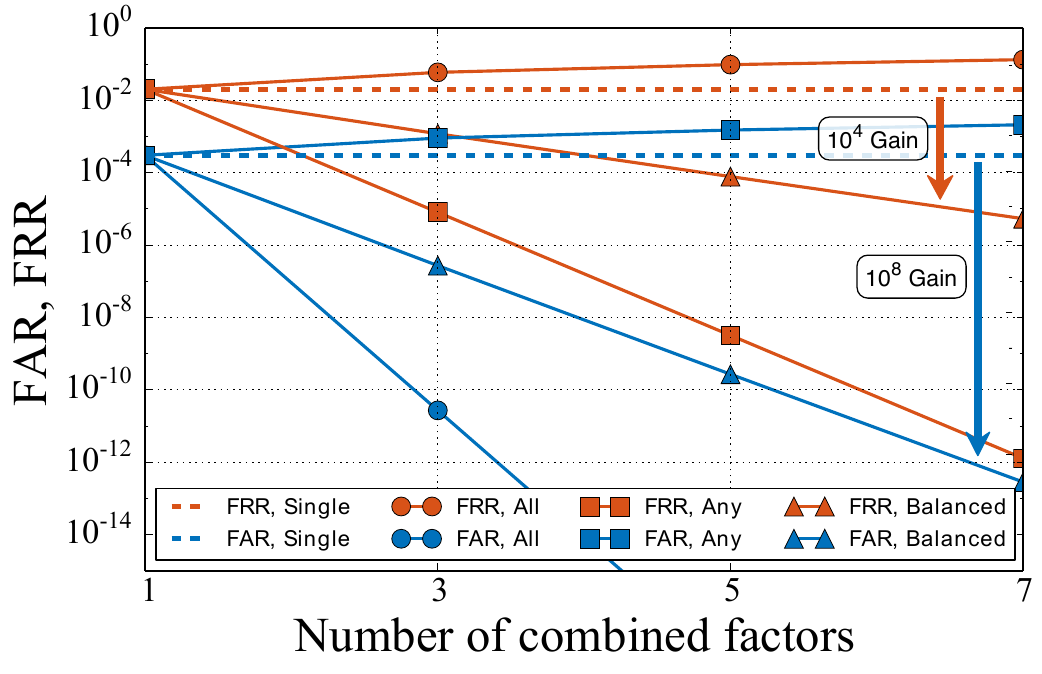}
		\caption{Comparing alternative factor combining approaches.}
		\label{fig:far}
	\end{figure}

	Observing Fig.~\ref{fig:far}, the \emph{\underline{All}} approach has the lowest FAR, thus yielding the best security level. However, its FRR is higher than with other approaches, by reaching over $12\%$ with~$7$ independent factors combined. Hence, the usability of \emph{\underline{All}} approach remains low, which makes it non-applicable in the A-IoT context. Further, we notice the opposite trend for the \emph{\underline{Any}} approach that increases the FAR value at the expense of much better FRR. Therefore, \emph{\underline{Any}} solution is not applicable either. Consequently, none of the trivial MFA combinations are directly usable in the challenging A-IoT scenarios.
	
	In contrast, a more intelligent \emph{\underline{Balanced}} approach -- ``A~user should successfully pass most of the checks to receive access''~-- constitutes a viable compromise between security and usability, by decreasing both FAR and FRR indicators. The quantitative gains highly depend on the input parameters and reach $10^4$ vs. $10^8$ when $7$ factors are combined. This example also highlights the importance of a threshold value selection, since incorrect combining may often result in rapid system performance degradation~\cite{ometov2018multi}. The same holds true for any other values of FAR/FRR, even though they may actually vary for different~factors.
	
	\vspace{-0.3cm}
	\subsection{Means to Evaluate}
	\textcolor{black}{Given that A-IoT scenarios are highly heterogeneous, the results delivered by the individual devices should not lead to blind acceptance/rejection decisions.} Instead, additional data must be considered when comparing the output of the authentication function against a threshold value.
	
	\begin{figure*}[!ht]
		\centering
		\includegraphics[width=1.65\columnwidth]{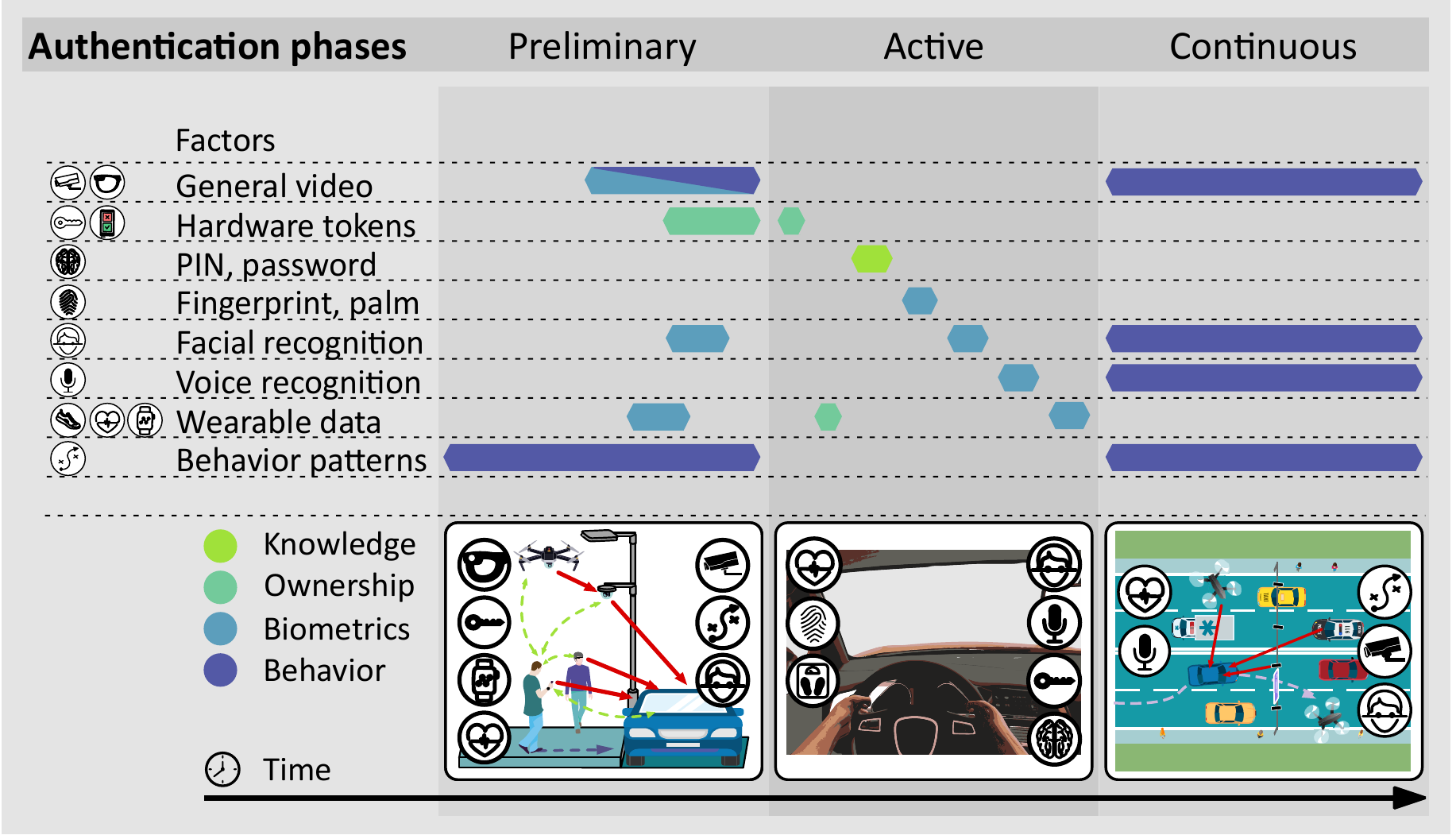}
		\caption{Considered phases of time-separated MFA for A-IoT.}
		\label{fig:potencial_util}
	\end{figure*}
	\subsubsection{Binary decision}
	The first and foremost sub-factor to be considered is binary decisions delivered by the individual devices.
	
	\subsubsection{Vendor-specific metrics}
	The second sub-factor is the level of accuracy, which is directly related to FAR/FRR parameters. For example, the data collected with cameras by various vendors may deliver different probabilities during a facial recognition event for the same user.
	
	\subsubsection{Level of trust}
	Many factors may impact user and device trust. Here, trust in the ``owned'' devices (e.g., built-in cameras) should be valued higher than that in external equipment. Further, historically familiar devices may have higher trust levels than stranger equipment, see the paradigm of Social~IoT~\cite{atzori2014smart}. A significant benefit may be made available by utilizing social networks, since the devices owned by a friend or a colleague may also be considered as more trustworthy.
	
	The set of selected sub-factors can significantly affect the operation of the authentication solution. However, the above three factors are relatively stable -- the overall changes in the A-IoT system from these perspectives are not as abrupt and thus could be determined in advance. Conversely, the authentication system designer should be provided with a higher level of flexibility for a given application. This could be achieved by adding another dimension -- specific factor weight per application (or even per user). 
	
	Accordingly, the general authentication function is to be considered as $\sum{\delta_i\mu_i\tau_i\varphi_i}>T,$    where $i$ is the factor number, $\delta_i$ is a binary decision, $\mu_i$ is the accuracy level provided by the vendor, $\tau_i$ is the trust level to the selected source, $\varphi_i$ is the factor weight, and $T$ is the system-wide threshold set by the designer. \textcolor{black}{Hence, the system may be adjusted per device, while the ultimate decision can be made flexible based on statistical analysis and machine learning techniques.} Finally, the use of various factors consumes different amounts of time, see Table~\ref{tab:comp}. 
	\vspace{-0.4cm}
	\subsection{Means to Evolve}
	
	The conventional ICT systems typically exploit a single-stage authentication method, such that the user is either granted or denied access as the result of authentication. In contrast, the more stringent time constraints of A-IoT authentication dictate the need to complement the main authentication phase with additional checks that happen before and/or after it. Here, the considered MFA solution may benefit from a range of sensing devices widely deployed in Smart Cities as well as exploit the very nature of the human interaction with the A-IoT system. Therefore, the overall authentication process can be divided into several phases and, consequently, the level of trust to the user begins to evolve in time. 
	
	\subsubsection{Pre-authentication phase}
	This phase is the most dynamic and unpredictable as a person `approaches' the target vehicle. Here, the surrounding environment plays a crucial role by providing with additional information. The only option during this phase is to utilize passive authentication strategies, i.e., `observe' the user biometrics/behavior that could be delivered by user-worn wearables, user-carried deceives, and other vehicles/infrastructure in proximity.
	
	\subsubsection{Active authentication phase}
	The most conventional phase relies upon active interaction. Hence, the user provides relevant input to the system directly. The most suitable authentication methods are knowledge- and biometrics-based.
	
	\subsubsection{Continuous (post) authentication phase}
	Another key part of the envisioned A-IoT authentication process is continuous monitoring of the fact that the user remains legitimate to operate the system even after the previous phases are completed successfully~\cite{8291131}. Monitoring and analyzing the subject by the smart vehicle, infrastructure, and other cars become a preferred option. Consider a case where the driver has provided all of the tokens, passed all of the biometric tests but faced a seizure during a highway trip. In this case, the vehicle may automatically overtake the control, connect to the neighboring cars, and safely stop by the wayside. As an example, recent works confirm that it is necessary to monitor the driver for just under 2.5 minutes in order to validate the behavior with 95\% accuracy~\cite{burton2016driver}.
	
	\vspace{-0.3cm}
	\section{Ecosystem of MFA-Powered A-IoT}
	\label{sec:challenges}
	
	The previous section summarized the underlying design principles of the MFA solutions in A-IoT, at large. However, even if these principles are followed, further development and mass adoption of MFA-powered A-IoT systems should be considered in perspective. This section brings the community's attention to the most significant questions to be answered in this context. 
	
	\subsubsection{How to weigh factors?}
	While the MFA concept offers sufficient flexibility to adapt the authentication system to a wide range of possible scenarios, the choice of particular numerical weights and threshold values requires an extensive study, which needs to carefully balance the FAR and FRR values of the resulting system depending on the target use case. The system should also be made reconfigurable, such that its internal parameters are updated appropriately whenever an A-IoT device is, e.g., sold to another person with different~attributes.
	
	\subsubsection{How to adapt decisions?}
	\textcolor{black}{Another critical challenge is dynamic system adaptation in relation to a number of factors involved in the authentication process.} For instance, recognition based on a video camera may be unavailable at nighttime or in bad weather. Hence, the decision function should dynamically adjust the weights of the factors that are available during the authentication process based on contextual data. This task is much more challenging as compared to conventional single- and two-factor authentication with only a few static factors~involved.
	
	\subsubsection{How to earn user trust?}
	The next question is related to making a legitimate user trust the system in its operations. For example, the user had a video surveillance camera at the parking near home, which contributed 20\% to the overall authentication process, while the threshold was configured to grant access. Then, the user moved the car to another address and cannot open it anymore without an additional weight from the infrastructure, since there is no external camera nearby to participate in the authentication process.     \textcolor{black}{Hence, it is crucial that the decision-making process be at least partially transparent to the user.}
	
	
	\subsubsection{How to receive assistance?}
	The A-IoT framework involves not only in-built authentication factors but also data from proximate sources.     \textcolor{black}{Therefore, a question remains of how secure and trusted such assistance from the neighboring devices could be.} Our illustrative example considered above receives additional data from the wearable devices owned by the human user; the camera mounted on a lamp post; a surveillance drone patrolling the street, etc. Hence, designing secure and reliable methods to deliver the sensitive authentication data from these dissimilar Smart City devices to the target A-IoT system -- while not compromising the user privacy for third-party entities -- is an open problem.
	
	\subsubsection{How to delegate A-IoT devices?}
	\textcolor{black}{Users tend to share their devices both privately~(family) and publicly~(car rent).}     \textcolor{black}{However, secure collective delegation of use is not straightforward for the A-IoT systems.}  \textcolor{black}{Conventional landing of a physical token may not be a sufficient option anymore, since it does not necessarily verify the right to operate the A-IoT device.} From the A-IoT platform perspective, most of the factors related to its temporary user.	
	
	\vspace{-0.3cm}
	\section{Conclusion and Standardization Aspects}
	Reliable and secure human authentication by various smart devices is one of the key drivers in the Advanced IoT era. 
	
	\textcolor{black}{From the standardization perspective, there is a number of regional specifications and recommendations related to multi-factor authentication. However, most of them are still in their early development phases. For example, Payment Card Industry Security Standards Council provides recommendations for the MFA system implementation\footnote{``PCI Security Standards Council: Guidance for Multi-Factor Authentication,'' \url{https://www.pcisecuritystandards.org/pdfs/Multi-Factor-Authentication-Guidance-v1.pdf} [Accessed November 2018]} and also partially touches upon the MFA-related topic in terms of the requirements related to the utilization of MFA for card payments in PCI~DSS~v3.2. National Institute of Standards and Technology~(NIST) provides MFA-related guidelines\footnote{``NIST Special Publications Portal,'' \url{https://www.nist.gov/publications} [Accessed November 2018]} in NIST Special Publications 800-63B and 800-63C with a detailed overview of the technical requirements for federal agencies implementing digital identity in the US. Overall, these documents support the discussion provided in this work. However, so far there is no unified standard for the MFA system developers to follow.}
	
	In this article, we reviewed the existing research challenges and possible enablers for user authentication within the A-IoT ecosystem. We introduced a concept of multi-factor authentication for A-IoT as an attractive alternative to existing single-factor solutions with limited potential.     \textcolor{black}{The fundamental design principles of MFA were highlighted by providing useful insights into facilitation of future MFA applications for the A-IoT.} Finally, key open questions related to the development, practical implementation, and adoption of MFA for diverse A-IoT systems were discussed together with potential use cases, thus laying the foundation for further research in this emerging~area.

	\vspace{-5mm}
	\bibliographystyle{ieeetr}
	\bibliography{refs}

\begin{thebibliography}{10}

\bibitem{extra1}
Y.~Yan {\em et~al.}, ``An efficient security protocol for advanced metering
  infrastructure in smart grid,'' {\em IEEE Network}, vol.~27, pp.~64--71, July
  2013.

\bibitem{cisco2017global}
{VNI Cisco}, ``Cisco visual networking index: Forecast and trends 2017--2022.''
  White Paper, 2018.

\bibitem{lin2017survey}
{J. Lin \textit{et al.}}, ``{A survey on Internet of Things: Architecture,
  enabling technologies, security and privacy, and applications},'' {\em IEEE
  Internet of Things J.}, vol.~4, no.~5, pp.~1125--1142, 2017.

\bibitem{shamir2017summary}
A.~Shamir, A.~Biryukov, and L.~P. Perrin, ``{Summary of an Open Discussion on
  IoT and Lightweight Cryptography},'' in {\em Proc. Early Sym. Crypto W.},
  University of Luxembourg, 2017.

\bibitem{guo2017survey}
J.~Guo, R.~Chen, and J.~J. Tsai, ``{A survey of trust computation models for
  service management in Internet of Things systems},'' {\em J. Comp. Comm.},
  vol.~97, pp.~1--14, 2017.

\bibitem{zhou2017security}
{J. Zhou \textit{et al}}, ``{Security and Privacy for Cloud-Based IoT:
  Challenges, Countermeasures, and Future Directions},'' {\em IEEE Comm. Mag.},
  pp.~26--33, January 2017.

\bibitem{joy2018internet}
J.~Joy, V.~Rabsatt, and M.~Gerla, ``{Internet of Vehicles: Enabling safe,
  secure, and private vehicular crowdsourcing},'' {\em Internet Technology
  Lett.}, 2018.

\bibitem{pascal}
S.~Tangade, S.~S. Manvi, and P.~Lorenz, ``Decentralized and scalable
  privacy-preserving authentication scheme in {VANET}s,'' {\em IEEE Trans. on
  Vehic. Tech.}, 2018.

\bibitem{corneanu2016survey}
{C. A. Corneanu \textit{et al.}}, ``{Survey on RGB, 3D, Thermal, and Multimodal
  Approaches for Facial Expression Recognition: History, Trends, and
  Affect-Related Applications},'' {\em IEEE Trans. on Pattern Analysis and
  Machine Intelligence}, vol.~38, no.~8, pp.~1548--1568, 2016.

\bibitem{ometov2018multi}
{A. Ometov \textit{et al.}}, ``{Multi-Factor Authentication: A Survey},'' {\em
  Cryptography}, vol.~2, no.~1, p.~1, 2018.

\bibitem{katsini2016security}
{C. Katsini \textit{et al.}}, ``{Security and Usability in Knowledge-based User
  Authentication: A Review},'' in {\em Proc. 20th Pan-Hellenic Conf. on
  Informatics}, p.~63, ACM, 2016.

\bibitem{benaliouche2014comparative}
H.~Benaliouche and M.~Touahria, ``{Comparative Study of Multimodal Biometric
  Recognition by Fusion of Iris and Fingerprint},'' {\em Scientific World J.},
  2014.

\bibitem{atzori2014smart}
L.~Atzori, A.~Iera, and G.~Morabito, ``{From ``smart objects'' to ``social
  objects'': The next evolutionary step of the Internet of Things},'' {\em IEEE
  Comm. Mag.}, vol.~52, no.~1, pp.~97--105, 2014.

\bibitem{8291131}
{K. H. Yeh \textit{et al.}}, ``{I Walk, Therefore I Am: Continuous User
  Authentication with Plantar Biometrics},'' {\em IEEE Comm. Mag.}, vol.~56,
  pp.~150--157, Feb 2018.

\bibitem{burton2016driver}
{A. Burton \textit{et al.}}, ``{Driver identification and authentication with
  active behavior modeling},'' in {\em Proc. 12th Int.'l Conf. on Network and
  Service Management}, pp.~388--393, IEEE, 2016.

\end{thebibliography}

\end{document}